\documentstyle[12pt]{article}

\newcommand{\be}{\begin{equation}}
\newcommand{\ee}{\end{equation}}
\newcommand{\bea}{\begin{eqnarray}}
\newcommand{\eea}{\end{eqnarray}}

\begin{document}
\begin{titlepage}

%\flushright{AEI-2001-099 }

\vspace{1in}

\begin{center}
\Large
{\bf SYMMETRIES OF AXION-DILATON STRING COSMOLOGY }

\vspace{1in}

\normalsize

\large{  Jnanadeva  Maharana\\
E-mail maharana$@$iopb.res.in 
  }

\normalsize
\vspace{.5in}

 {\em Institute of Physics \\
Bhubaneswar - 751005 \\
India  \\ }

\end{center}

\vspace{1in}

\baselineskip=24pt
\begin{abstract}
The axion-dilaton string effective action is expressed in Einstein
frame metric in a manifestly  S-duality invariant form. It is shown
that the moduli can be redefined to describe surface of a 
$(2+1)$-dimensional pseudosphere. The classical cosmological solutions 
of axion-dilaton
are revisited. The Wheeler De Witt equation for the system is exactly solved
and complete set of eigenfunctions are presented. The wave function
factorizes and the one depending on the moduli is obtained by appealing
to the underlying S-duality symmetry. Axion and dilaton parametrize
${{SU(1,1)}\over {U(1)}}$, the S-duality group, and the Hamiltonian is 
expressed as a sum of the Ricci scalar and the Casimir of $SU(1,1)$. Therefore,
the complete set of wave functions, depending on the moduli, is obtained from
group theory technique. The evidence for the existence 
of a W-infinity algebra in axion-dilaton
cosmology is presented and the origin  of the algebra is primarily 
 due to high
degree of degeneracies in the wave functions. It is qualitatively argued
 that axion-dilaton quantum cosmology exhibits chaotic
behavior in the semiclassical limit.

\end{abstract}

\vspace{.7in}
 
\end{titlepage}

\section{Introduction }

It is well known that string theories possess rich symmetry structures. 
These symmetries have played central role in our understandings of
various facets of string theories. Notable among the symmetries are dualities
which have enabled us to explore string dynamics in diverse dimensions
and have been guiding principles to seek an underlying fundamental theory in
order to unify the five string theories \cite{john1}. The predictions of target
space duality, T-duality, could be to tested in the frame work of
 perturbative
string theory. On the other hand, the consequences of 
strong/weak coupling duality, the S-duality, 
are naturally  nonperturbative in character. \\
One of the common attributes of the five perturbatively consistent string
theories is that gravity is naturally incorporated in all of them and 
therefore, we can seek answers to deep questions in quantum gravity within
the frame work of string theory. Indeed, string theory has provided 
deeper understandings about stringy black holes in the sense that the
Bekenstein-Hawking entropy relation can be derived for a special class
of black holes from the counting of microscopic states.\\
It is natural to explore the cosmological scenario from the perspectives
of string theory. We expect that string theory will play important role in
studying the evolution of the Universe in early epochs and shed lights on
the mechanism of inflation. It is hoped that string theory will provide
resolution of the initial singularity problem in cosmology.
There have been considerable amount of activity in order to investigate
various aspects of string cosmology during the past decade \cite{rev1,rev2}. 
In this context, symmetries too have played a key role in formulating
stringy cosmological models. The pre-big bang (PBB) \cite{pbb}
 proposal has received a lot of attention since it incorporates stringy
symmetries in order to provide a novel mechanism for inflation.\\ 
It is worth while to recall the role of dilaton, $\phi$,  in string theory.
The dilaton  appears as a massless excitation in string theory 
along with other massless excitations of the  spectrum. 
An  important point is that the exponential of the VEV of dilaton is the
coupling constant in string theory and in turn it  
determines the Newton's constant, the gauge coupling constants and Yukawa
couplings of low energy effective actions derived to describe models of
particle physics. 
Moreover, when one envisages cosmological situation, the 
dilaton naturally appears in the effective action along with graviton. 
In fact the graviton-dilaton pair play a crucial role for the existence
of scale factor duality (SFD) \cite{gab1} in string cosmology and the
analog of this symmetry  does not exist in pure Einstein gravity.
 Moreover, it is 
well known that the the pair dilaton, $\phi$, and axion, $\chi$, parametrize
the S-duality group \cite{crem,sdu}
${SL(2,R)\over {U(1)}} \sim {SU(1,1)\over {U(1)}}$.
The axion, as a weakly coupled pseudoscalar boson, is incorporated
in the standard model of particle physics and is expected to play a vital role
in cosmology.  
 Therefore, it is interesting to explore various
aspects of axion-dilaton cosmology in the string theory.\\
This article is a continuation of our  study of classical and quantum aspects of
axion-dilaton string cosmology. Recently, we have presented some of our 
preliminary results in a short communication \cite{j1}.
We briefly discuss known classical solutions
for the system and explore further the quantum cosmological solutions. The
S-duality symmetry plays a central role in our investigation. At the classical
level, we construct the generators of the symmetry and present the algebra
of the generators. The problem may be conveniently studied by a suitable
change of variable where the axion dilaton moduli space is described as the
surface of a $2+1$ dimensional pseudosphere. In the cosmological scenario,
the moduli depend on cosmic time coordinate and the classical Hamiltonian
constraint reflects invariance of the action under time reparametrization.
In quantum cosmology, the Hamiltonian constraint translates to an operator
condition, the Wheeler De Witt equation \cite{wdw}.\\
We present a class of solutions to WDW equation for the case at hand. These
solutions are new, were missed in earlier investigations \cite{mmp,rev3}
 and are obtained from purely group theoretic considerations
for the moduli describing surface of the pseudosphere. 
It is found that the wave function is highly
degenerate. Furthermore, we explore the possibility of solving WDW equation
for the $\chi - \phi$ system directly. It is shown that, with a well known
redefinition of these fields, the problem can be mapped to that of 
motion on the Poincar\'e upper half plane. The classical problem, in the
absence of coupling to gravity has been studied quite well. The solution
to classical equations of motion, for axion-dilaton system coupled to gravity,
are well known. In the context of quantum
mechanics, it turns out that separation of variable for the wave function of
the Universe is possible when we write down the WDW equation. Then
one has to solve a differential equation involving the scale factor and
another one which is the analog of Schr\"odinger equation 
in the Poincar\'e upper 
half plane. Since the problem of motion
on surface of a pseudosphere and that of Poincar\'e upper half plane are
related \cite{balv}, it is not surprising that 
the wave function derived in the latter
formulation is also highly degenerate. One of the main difference in the 
characteristics of the wave function derived for motion of pseudosphere is that
we seek a simultaneous eigenfunction of the Casimir operator and the only
compact generator of $SU(1,1)$. In case of the Poincar\'e upper half
plane, we choose solutions such that the wave function can be
expressed as a product of a plane wave in the axion field and a function 
of the dilaton. Obviously, the plane wave part has continuous eigenvalue
and is related to an eigenfunction of a noncompact generator of the
S-duality group. There is some advantage in solving the quantum equation in
terms of coordinates defining the Poinca\'re upper half plane. Here, we can
directly study the behaviour of the wave function in the strong coupling limit
which is not so transparent when we solve 
WDW equation in terms of the coordinates
of the pseudosphere. On the other hand, some of the symmetry properties are
more directly brought out in formulating the problem with coordinates of the 
pseudosphere.\\
The WDW equation is  a second order differential which will have
infinite number of solutions in general. Therefore, it is necessary to
supplement the solutions with boundary conditions. When we seek solutions to 
wave equations in quantum mechanics, the boundary conditions are imposed by
taking physical considerations into account from out side. In contrast, when one
derives the wave function of the Universe, the boundary conditions are
generally considered  as additional principles.
The well known boundary conditions are the Hartle-Hawking no boundary proposal
\cite{hh} and the  tunneling scenario advanced by Linde and Vilenkin 
\cite{lin,vil}.\\
In the context of string cosmology, it is necessary to envisage the
boundary conditions in an appropriate perspective. Here, dilaton always
accompanies graviton and it is also the coupling constant as alluded to
earlier. Thus, when the boundary condition on the wave function of the
Universe is considered, this aspect should be kept in mind. In the
PBB scenario, the classical solution is such that initially, the
Universe evolves from low curvature, low temperature phase and is in weak the 
coupling regime. Subsequently,  it undergoes accelerated expansion. When we
consider quantum string cosmological equation in the frame work of
PBB proposal, the boundary condition on the wave function is not introduced
from outside \cite{rev2,gmvq}. 
The PBB mechanism already dictates which boundary condition
is to be respected by the wave function. Furthermore, in the string frame,
the Hamiltonian manifestly is invariant under the scale factor duality and
therefore, the quantum Hamiltonian is determined so as to respect 
this symmetry. Note that in this case the string symmetries impose additional 
conditions on the wave functions of WDW equation \cite{gmvq}. Therefore,
one might contemplate that a third  of boundary condition is available
as an option in quantum string cosmology.\\
There is a general trend in string theory (in the sense of
string effective action) that as we envisage theories in lower spacetime
dimensions, there is enhancement of symmetries. Let us very briefly recapitulate
some of the important features of string effective action in this perspective. 
We may recall a generic feature of the toroidal
compactification of string effective action from critical, $10$, dimension
to lower dimension. If we consider heterotic string as an illustrative example,
we note that the four dimensional theory possesses, not only the T-duality
group $O(6,22)$, but also it is endowed with S-duality symmetry (at the level of
equations of motion). When one compactifies the theory on seven torus to 
go to three dimensions the group in $O(8,24)$ which contains both T-duality
group, $O(7,23)$ as well as the S-duality group, $SL(2,R)$ 
\cite{duff}. Furthermore, the
the theory in two spacetime dimensions is known to admit 
affine algebras \cite{mss}.
Thus, it might be interesting to explore whether the theory depending on one
coordinate, in the present context the cosmological scenario, 
contains an underlying symmetry other than the S-duality which is manifest
in the problem under consideration. We   have discovered a $w_{\infty}$
algebra whose generators are constructed from the S-duality group, $SU(1,1)
\sim SL(2,R)$
and they act on the space of solutions of the WDW equation \cite{j1}.\\ 
The plan of the paper is as follows: In Section 2, we present the
tree level string effective action for graviton, axion and dilaton in
the Einstein frame. We rewrite the action in a manifestly S-duality
invariant form and the we identify the group  to be $SU(1,1)$ and 
discuss the invariance properties of the action. The conserved currents,
associated with the symmetry transformations, are constructed from the
standard N\"other procedure. We study the cosmological scenario in Section 3.
First, the classical cosmological axion-dilaton solutions are
briefly discussed. Next, the Hamiltonian constraint is derived and the
Wheeler De Witt equation is obtained. 
It is argued that the wave function could be expressed in the factorized form
separating into function of the scale factor $a$ and that of the moduli.
The WDW equation is first solved in the
Lorentzian coordinates representation of $(2+1)$-dimensional the pseudosphere. 
The eigenfunctions
for the moduli are presented from the group theoretic considerations and
their properties are elucidated. Next, wave function depending the scale factor
is obtained. The solutions to WDW equation is also given when the moduli
is expressed in the Poincar\'e coordinates defining the moduli space
to be the upper half plane. This choice of coordinate system has the
advantage that the strong coupling limit of the wave function is rather
transparent and it is easy to exhibit the semiclassical limit of the
theory in simple way.  Section 4 contains  discussion of
the W-infinity algebras in axion-dilaton cosmology which is a more
detailed analysis of our preliminary conjecture \cite{j1}.
In Section 5, we qualitatively explore the phenomena of chaos in axion-dilaton
quantum cosmology. The last Section is devoted to  summary and discussion
 of our results. The Appendix contains some relevant formulas for the
$SU(1,1)$ group.  
 
\section{Effective Action and S-duality Symmetry}
Let us consider a four dimensional string effective action in the presence of
 axion and dilaton.
\bea
\label{axd}
S_4=\int d^4x{\sqrt {-g}}\left(R-{1\over 2}(\partial \phi)^2 -{1\over 2}
e^{2\phi}(\partial \chi)^2 \right) \eea
Here $R$ is the scalar curvature, computed from the Einstein frame metric
$g_{\mu\nu}$, $\sqrt {-g}$ is its determinant and the other two terms 
correspond to kinetic energy terms of dilaton and axion respectively. This
action may be thought of as the one derived from ten dimensional type IIB
effective \cite{jm2b} action where rest of the background fields are set to
 zero. Alternatively, we could derive this action from NS-NS sector of a string
theory where the field strength of $B_{\mu\nu}$, $H^{\mu\nu\lambda}= e^{2\phi}
\epsilon ^{\mu\nu\lambda\rho}\partial _{\rho}\chi$, is dualized to introduce the
axion. It is more convenient to express the action in a manifestly S-duality
invariant form as given below
\bea
\label{manisd}
S_4=\int d^4x{\sqrt{-g}}\left(R+{1\over 4}{\rm Tr}[\partial _{\mu}
{\bf V}^{-1}(x)
\partial ^{\mu}{\bf V}(x)] \right) \eea
where the $2\times 2$ matrix $\bf V$ is given by
\bea
\label{vmatrix}
{\bf V} ={1\over 2}\pmatrix {A+B & 2B\chi +i(A-B) \cr 2B\chi -i(A-B) & A+B \cr}
\eea
The elements of the $2\times 2$
matrix $\bf V$ are defines as: $A=e^{-\phi}+\chi ^2e^{\phi}
$ and  $B=e^{\phi}$. Note that ${\bf V} \in SU(1,1)$ and satisfies
${\bf V}^{-1}=\sigma _3{\bf V}^{\dagger}\sigma _3$ and  
$\sigma _i ,i=1,2,3$
are the Pauli matrices. The action (\ref{manisd}) is invariant under
\bea
{\bf V}\rightarrow \Omega ^{\dagger}{\bf V}\Omega, ~~~g_{\mu\nu}\rightarrow
g_{\mu\nu} ~~~{\rm and}~~~\Omega ^{\dagger}\sigma _3 \Omega =\sigma _3,
\eea
where $\Omega \in SU(1,1)$ and $\sigma _3$ is the metric which is required
to be invariant under the group transformations. We recall that any $2\times 2
$ matrix ${\bf U} \in SU(1,1)$, in general is be spacetime dependent,
 can be expressed as
\be
{\bf U}={\bf 1}U_0+\sigma _1 U_1+\sigma _2 U_2+i\sigma _3 U_3
\ee
and the spacetime dependent coefficients satisfy the constraint
\be
U_0^2+U_3^2-U_1^2-U_2^2=1
\ee
Note from the structure of the $\bf V$ matrix (\ref{vmatrix})
that, if we expand it in terms
of unit matrix and $\sigma _i$, the coefficient of $\sigma _3$ 
turns out to be zero;
therefore, the constraint satisfied by the coefficients $v_i,~ i=0,1,2$ is
\be 
\label{constraint}
v_0^2-v_1^2-v_2^2=1 \ee
Therefore, the moduli define surface of a 
$(2+1)$-dimensional psuedosphere as mentioned earlier.
We can read off the coefficients from (\ref{vmatrix})
\be v_0={1\over 2}(A+B),~~v_1=B\chi ,~~ v_2={1\over 2}(B-A)  \ee
and it is easy to check that they satisfy the requisite constraint 
(\ref{constraint}).\\
We may write the action (\ref{manisd}) in terms of the moduli as
\be
\label{vaction}
S=\int d^4x {\sqrt{-g}}\left( R+{1\over 2}g^{\mu\nu}\eta ^{ij}\partial _{\mu}
v_i \partial _{\nu}v_j \right)
\ee
The metric in this moduli space 
is $\eta ^{ij} ={\rm diag}(1,-1,-1)$ with $i,j=0,1,2$ 
and the constraint 
(\ref{constraint}) reads $\eta ^{ij}v_iv_j=1$.\\
The generators of $SU(1,1)$, denoted by $J_1 , ~J_2 ,~{\rm and}~J_3$ satisfy the
commutation relation
\be
\label{cr}
[J_1 ,J_2]=-iJ_3 ,~~[J_2 ,J_3]=iJ_1 ,~~ [J_3 ,J_1]=iJ_2
\ee
The set of $2\times 2$ matrices satisfying the $SU(1,1)$ algebra are
\bea
J_1={1\over 2}\pmatrix {0 & i \cr i & 0 \cr} , ~~J_2={1\over 2}\pmatrix {0 & 1 \cr
-1 & 0 \cr } , ~~~{\rm and}~~~J_3= {1\over 2}\pmatrix {1 & 0 \cr 0 &-1 \cr}
 \eea
As is well known, for a noncompact group like $SU(1,1)$, it is not possible
to choose all generators as $2\times 2$ Hermitian matrices. Now let us
consider an infinitesimal $SU(1,1)$ transformations on the moduli $v_i$. We can
express an $SU(1,1)$ transformation as 
\be
\Omega =e^{i\alpha _kJ_k} \sim 1+i\alpha _kJ_k \ee
for the infinitesimal parameters $\{ \alpha _i \}$. Thus the variation of the
$\bf V$ matrix $\delta \bf V = \Omega ^{\dagger} {\bf V}\Omega -{\bf V}$
is expressed as
\be
\delta {\bf V}={\bf 1}\delta v_0+\sigma _1 \delta v_1 + \sigma _2 \delta v_2
\ee
and the variations of $v_i$ are given by
\be
\delta v_0=-(\alpha _1v_1+\alpha _2v_2),~~\delta v_1=-(\alpha _1v_0+
\alpha _3v_2), ~~\delta v_2=\alpha _3v_1-\alpha _2v_0 \ee
We remind the reader that the Einstein frame metric remains invariant
under S-duality transformation and it is rather straight forward to construct
the three conserved currents from the action (\ref{vaction}).
\be
{\cal J}_1^{\mu}={\sqrt {-g}}(\partial ^{\mu}v_1v_0 -\partial ^{\mu}v_0v_1) \ee
\be
{\cal J}_2^{\mu}={\sqrt {-g}}(\partial ^{\mu}v_2v_0-\partial ^{\mu}v_0v_2) \ee
\be
{\cal J}_3^{\mu}={\sqrt{-g}}(\partial ^{\mu}v_1v_2 -\partial ^{\mu}v_2v_1) \ee
The charge densities can be expressed as
\bea
{\cal Q}_1=(p_1v_0+p_0v_1),~~{\cal Q}_2=(p_2v_0+p_0v_2),~~{\cal Q}_3=(p_1v_2
-p_2v_1) \eea
where the canonical momenta are defined from the Lagrangian given by
(\ref{vaction}). We mention in passing that the generators ${\cal Q}_1$
and ${\cal Q}_2$ are noncompact and they are like Lorentz generators on
the pseudosphere; whereas, ${\cal Q}_3$ is compact and generates rotation
on  $v_1 - v_2$ plane. The charge densities satisfy $SU(1,1)$ algebra (\ref{cr})
 when we compute their Poisson brackets and the factor of '$i$' is recovered
when the naive commutation relations are carried out ignoring the subtleties
that arise in product of operators.  
\\
It is also useful to explore how the $SU(1,1)$ algebra emerges starting
from action (\ref{axd}). Let us define a complex scalar field
\be
\label{lamd}
\lambda =\lambda _1+i\lambda _2 ,~~{\rm where}~~\lambda _1=\chi ~~{\rm and} ~~
\lambda _2=e^{-\phi} 
\ee
Then (\ref{axd}) is expressed as
\bea
S_{\lambda}=\int d^4x{\sqrt{-g}}\left(R-{1\over {2{\lambda _2}^2}}g^{\mu\nu}
\left[\partial _{\mu}\lambda _1\partial _{\nu}\lambda _1 +\partial _{\mu}
\lambda _2\partial _{\nu}\lambda _2\right]\right)
\eea
Let us consider following set of transformations which leave the Einstein
metric invariant and act only on $\lambda _1 ~{\rm and} ~\lambda _2$.\\
(i) Dilation:
$ \delta \lambda _1=\epsilon _1\lambda _1 , ~~\delta \lambda _2=\epsilon _1
\lambda _2 $.\\
(ii) Translation along $\lambda _1$: $\delta \lambda _1=\epsilon _2 , \delta
\lambda _2=0$.\\
(iii) Nonlinear rotation: $\delta \lambda _1=\epsilon _3(\lambda _2^2
-\lambda _1^2),~~{\rm and }~~\delta \lambda _2=-2\epsilon _3\lambda _1
\lambda _2 $.\\
Where $\epsilon _i ,~i=1,2,3$ are constant parameters. We may derive the
conserved currents, following N\"other's prescriptions. The charge densities
are given by

\be
Q_1=\lambda _1p_{\lambda _1}+\lambda _2p_{\lambda _2} ,~~Q_2=p_{\lambda _1},
~~Q_3=p_{\lambda _1}(\lambda _2^2-\lambda _1^2)-2\lambda_1\lambda_2
p_{\lambda _2} \ee
It is easy to check that the Poisson bracket algebra of three charges obtained
by the spatial volume integration of the densities $Q_i$ close
on to algebra of the generators of the S-duality group. The following linear
combination of the three charges satisfy the algebra of generators given in
(\ref{cr})
\be
J_1={1\over 2}(q_1+q_2),~~J_2=q_1~~{\rm and}~~J_3={1\over 2}(q_2-q_3)
\ee
where the charges, $q_i=\int dVQ_i$.

%%%%%%%%%%%%%%%%%%%%%%%%%%%%%%%%%%%%%%%%%%%%%%%%%%%%%%%%%%%%%%%%%%%%%%%%%%%%%%%
%This is SECTION 3 of axion-dilaton string cosmology
%%%%%%%%%%%%%%%%%%%%%%%%%%%%%%%%%%%%%%%%%%%%%%%%%%%%%%%%%%%%%%%%%%%%%%%%%%%

\section{The Cosmological Scenario }
We presented the symmetries of the four dimensional effective action
with axion and dilaton as spacetime dependent matter fields. 
The generators of the S-duality
group, $SU(1,1)$ were constructed in terms of the axion and dilaton fields
as well as in terms of the new fields $v_i$ which describe surface of
the pseudusphere. In the cosmological case, the metric and the matter fields
are assumed to depend on cosmic time. Furthermore, we adopt homogeneous,
isotropic FRW metric 
\bea
\label{frw}
ds^2=-dt^2+a(t)^2\left({{dr^2}\over {1-kr^2}}+r^2d\Omega ^2 \right)
\eea
$a(t)$ is the scale factor and $t$ is cosmic time. 
The constant, $k$,  takes values $k=+1,0, -1$
which corresponds to  closed, flat
and open Universes respectively. The Einstein field equation,
used to solve for  the scale factor, is given by
\be
\label{einstein}
R_0^0-{1\over 2} R=^{(\phi)}T^0_0+^{(\chi)}T_0^0 \ee
where $R_0^0$ is the $0-0$ component of the Ricci tensor and the right hand 
side is the sum of the $0-0$ components of stress 
energy momentum tensor for dilaton and axion as is obvious from the notation.
The action, in minisuperspace,  is given by
\bea
S=\int dt (-6a{\dot a}^2 +6ka +{1\over 2} a^3{\dot {\phi}}^2 +{1\over 2}
e^{2\phi}a^3{\dot {\chi}}^2 )
\eea
This action is expressed in terms of the Einstein frame metric.
In the above equation, we have not explicitly written the total derivative
term which arises when we partially integrate the piece coming from
the scalar curvature. As a simple, illustrative example, let us look at the
classical solutions of field equations for the case $k=0$. \\
The other two equations corresponding to axion and dilaton evolutions are
\be
\label{axeq}
{\ddot {\chi}}+3H{\dot {\chi}}+2{\dot{\phi}}{\dot {\chi}}=0 \ee
\be
\label{dileq}
{\ddot {\phi}}+3H{\dot{\phi}}=e^{2\phi}{\dot{\chi}}^2 \ee
where $H={{\dot {a}}\over a}$, is the Hubble parameter. Notice that the axion
equation leads to a charge conservation law since
\be
{{d\over {dt}}}\left( ln~{\dot {\chi}}+ln{a^3}+2\phi \right)=0 \ee
which implies $a^3{\dot{\chi}}e^{2\phi}=\pm L$, L being a 
constant of motion. The two 
time evolution equations for axion and dilaton (\ref{axeq}) and
(\ref{dileq}) may be solved
more efficiently \cite{edcop} if one introduces a line element 
\be
d\xi ^2=d\phi ^2+e^{2\phi}d\chi ^2 \ee
leading to equation of motion
\be
\label{friction}
{\ddot {\xi}}+3H{\dot {\xi}}=0 \ee
Again leading to a conservation law $a^3{\dot {\xi}}=\pm K$, where $K$ is a 
positive constant.Then one could solve the Einstein-Friedmann equation by 
utilizing the conservation law of the $\xi$ variable. It suffices for
our purpose to note that 
\bea
\label{classol}
a \sim t^{{1\over 3}},~~~~ e^{\phi}\sim {L\over {2K}} 
\left(({t\over {t_0}})^{-{2\over {{\sqrt 3}}}} +({{t_0}\over t})^
{-{2\over {{\sqrt 3}}}} \right)~~ {\rm and}~~{\chi}\sim {\rm Const}
{K\over L}{1\over {{1+({t\over {t_0}})^{-{4\over{\sqrt 3}}}}}}  \eea 
Note that the solutions presented here are in terms of cosmic time, $t$ and
$t_0$ is a constant of integration, In ref. \cite{edcop} 
the solutions were obtained in conformal time frame.
Let us dwell on some of the features of the  solution (\ref{classol}). We note
that the scale factor $a(t) \rightarrow 0$ as $t \rightarrow 0$ which
corresponds to curvature singularity. At this point dilaton diverges;
in other words, as $t \rightarrow 0$ the coupling constant becomes strong.
However, the axion tends to a constant value in the $t \rightarrow 0$
limit. As is well known, it is not possible to avoid the curvature singularity
by a different choice of coordinate frame. The classical solution of
axion-dilaton system has been studies in detail in \cite{edcop} for the
cases $k=\pm 1, 0$ and it was found that dilaton approaches strong coupling
limit in all these cases. Note that
 in absence of the axion (i.e. K=0), the solution is obtained for 
graviton-dilaton system in the Einstein frame, where we see the presence
of two solutions for the scale factor and also there are two different
solutions for the dilaton.\\
Now we proceed to analyze quantum axion-dilaton cosmology in more detail.
In order to set up the Wheeler De Witt equation with manifest S-duality
symmetry and to facilitate the solutions of the quantum mechanical
equations, 
 it is more convenient
to deal with the moduli $\{v_i \}$. Note that the scalar curvature
is given by i.e. ${\sqrt{-g}}R=6(-a{\dot a}^2+ka)$ for our choice of FRW metric
and we have dropped a total derivative term here. The canonical
momentum associated with the scale factor becomes 
$P_a=-12{\dot a}a$. Consequently,
when we obtain the canonical Hamiltonian there will be 
unconventional numerical factor like $1\over {24}$  factor 
multiplying $P_a ^2$ which are inconvenient to keep track of. Therefore, as
is the accepted prescription, we rescale the metric and the moduli $v_i$
accordingly so that the cosmological action takes the following form
\be
\label{newact}
{\tilde S}={1\over 2}\int dt\left( -a{\dot a}^2+ka -a^3{\dot v}_i{\dot v}_j 
\eta ^{ij} \right) \ee
 Let us consider  $k=1$  as a generic case from now on; $k=0$ will be rather
simple and $k=-1$ can be dealt with as one solves $k=+1$ situation.
 We shall see that
the solution to the 'angular part' of the WDW equation involving axion
and dilaton is not affected by the choice of $k$. This part of the
WDW wave function will be determined from purely group theoretic
considerations. Therefore, some of our general conclusions
are not affected by this choice of $k$. First, we derive the Hamiltonian 
constraint corresponding to (\ref{newact})
\bea
\label{hamiltonian}
{\cal H}={1\over 2}\left({1\over a}P_a^2+a +{{1\over {a^3}}}\eta _{ij}P^i_vP^j_v
\right) =0.
\eea
where ${\cal H}$ is the canonical Hamiltonian derived from (\ref{newact}) and
the canonical momenta are defined as
\be
\label{momenta}
P_a={{\partial L}\over {\partial {\dot a}}},~~P^i={{\partial L}\over {{\partial
{{\dot {v}}_i}}}} \ee
We remark in passing that the action (\ref{newact}) is invariant under 
the $SU(1,1)$
transformations and so is the Hamiltonian (\ref{hamiltonian}). It is easy
to check that the Hamiltonian  commutes
with the generators of $SU(1,1)$ as well as with  the Casimir.\\
When we implement (\ref{hamiltonian}) as a quantum constraint, ${\cal H}$
is replaced by operator, $\hat {\cal H}$, which acting on the wave function
is required to give zero eigenvalue. We identify ${\hat P}_a=-i{{\partial}
\over {\partial a}}$ and ${\hat P}^i=-i{{\partial }\over {\partial {v_i}}}$.
 Note
that the first term in the expression for the classical (\ref{hamiltonian}) 
contains a product of the scale fact, $a$, and its canonical momentum. In
defining the quantum Hamiltonian, we encounter the well known ordering
ambiguity since product of two noncommuting operators appears. We adopt
the the prescription such that the resulting Hamiltonian 
 respects invariance under change of
 coordinates  in
minisuperspace \cite{swh}.  
The next important point to note that the last term in (\ref{hamiltonian}) is
related to the Casimir operator of $SU(1,1)$ and therefore ${\cal H}$
is S-duality invariant. We can recognize this if we
consider 'polar' coordinates
\be
\label{polar}
 v_0={\rm cosh}\alpha , ~~v_1={\rm sinh}
\alpha {\rm cos}\beta~~{\rm  and}~~ v_2={\rm sinh}\alpha {\rm sin}\beta \ee
These variables are also called `disc variables';
where  $\alpha$ is real
and $0\le \beta \le 4\pi$, since the magnetic quantum number takes both integer
and half integer values as we shall see later. Furthermore, axion and dilaton
can be expressed as
\be
\label{axdilrl}
\chi={{{\rm sinh}\alpha {\rm cos}\beta}\over {({\rm cosh}\alpha +{\rm sinh}
\alpha {\rm sin}\beta)}},~~e^{-\phi}={1\over {({\rm cosh}\alpha +{\rm sinh}
\alpha {\rm sin}\beta)}} \ee
The Casimir operator, in terms of the 'polar'  coordinate system
 is just the Laplace-Beltrami
operator given by
\be
\label{lb}
 {\hat C}= -{1\over {{\rm sinh}\alpha}}{{\partial}\over {\partial \alpha}}
{{\rm sinh \alpha}}
{{\partial}\over {\partial \alpha}}-{1\over {{\rm sinh}^2\alpha}}{{{\partial}^2}
\over {\partial {\cal \beta}^2}}
\ee
If we denote eigenvalue of the Casimir operator, $\hat C$, by $C$, then
the quantum Hamiltonian satisfies WDW equation i.e. ${\hat {\cal H}}\Psi =0$
which assumes the following form as a differential equation
\bea
\label{wdw}
\left({{{\partial}^2}\over {\partial a^2}}+{{\partial}\over{\partial a}}-a^2
+{1\over a^2}{\hat C}\right){\Psi}=0 .
\eea
We may express $\Psi$ as product of two functions: one depending on the
scale factor, $a$ and other depending on the two 'polar' coordinates:
$\Psi ={\cal U}(a)Y$. Here $Y$ is the eigenfunction of the Casimir operator
${\hat C}Y=j(j+1) Y$. We may choose the eigenfunction to be also simultaneous
eigenfunction one one of the generators of $SU(1,1)$ and here we choose
that operator to the compact generator, $J_3$. Thus the goal is to solve
for the relations
\bea
{\hat C}|j,m>=j(j+1)|j,m>,~ {\rm and } ~J_3|j,m>=m|j,m>
\eea
 Now the problem is reduced to identifying all unitary, infinite dimensional
representations of $SU(1,1)$ satisfying above requirements. We present below
the classifications of the representations of $SU(1,1)$ which are well
 known \cite{su11,suph1,suph2} and refer to readers to the 
Appendix for some more relevant
details.\\
The representations of $SU(1,1)$ mainly fall into three different categories:\\
(i) The discrete series $D^{\pm}_j$.\\
$D^+_j$ is the one for which $j=-{1\over 2}, -1, -{3\over 2}, -2...$. For a 
given $j$ value, the eigenvalue of $J_3$, $m$, is unbounded from above taking
values $m=-j, -j+1, -j+2,...$.\\
On the other hand, for the other discrete series $D^-_j$, j is negative, taking
integer and half integer values. Moreover, in this case, $m$ is unbounded from
below and therefore, $m=j, j-1, j-2, ....$. We adopt the convention where
$j~is~negative$. \\
The two sets of wave functions are related by a symmetry when $m\rightarrow -m$
which follows from the properties of the $D^j_{mm'}$ functions of $SU(1,1)$
group (see the Appendix for definition of these functions).\\
(ii) The continuous series are also of two types; $C^0_l$ and $C^{1\over 2}_l$
 and $j$ turns out to be complex
in this case:\\
\be
 j=-{1\over 2} + il, ~~ l>0,~~{\rm and ~real} \ee
For $C^0_l$ the eigenvalues of $J_3$ take positive and negative integer value
for each $j$ i.e. $m=0,\pm 1, \pm2,....$. On the other hand, $m=\pm {1\over 2},
\pm{3\over 2},....$ for the $C^{1\over 2}_l$ series.
Note that $j^*=-j-1$ and also that the eigenvalue of the Casimir operator is
real.\\
(iii) The supplementary series is defines when the value of $j$ lies in the
range $-{1\over 2}<j<0$ and the magnetic quantum number takes integer values
in this case: $m=0,\pm 1, \pm 2,...$.
There is a theorem due to Bargmann which states that any function defined
on $SU(1,1)$ can be expanded in terms of the set of functions belonging
to discrete and continuous series $\{D^{\pm}_j, C^O_l, C^{1\over 2}_l \}$ and
it is not necessary to include the functions belonging to the supplementary
series.\\
At this stage, it is worth while to discuss a few points before we derive
the explicit form of the wave function. The constrain equation for $v_i$'s i.e.
$v_0^2-v_i^2-v_2^2 =1$;
 describes, as mentioned earlier, a surface of constant negative curvature
in the moduli space. The constraint equation is for the moduli and $v_0$ has
nothing to do, we emphasize, with spacetime coordinates at all. It is 
just one of the three moduli. The equation defining the pseudosphere has
interesting geometrical structure and let us recall some of these features
\cite{balv}. The locus of points equidistant of the origin specifies a
hyperboloid of two sheets intersecting the $v_0$-axis at the point $v_0=\pm 1$.
These two points, in analogy with the $S^2$ are called poles. Due to the
nature of this geometry, each sheet models an infinite space-like surface
and this surface has no boundary. Moreover, the geodesics are intersection
of hyperboloid sheet with planes through the origin.\\
Another comment is in order here. We have expressed the wave function of
the WDW equation (\ref{wdw}) in the factorized form and thus one of our goal is
to solve the wave equation on the pseudosphere. It is well known from earlier
studies \cite{grosch} that the solution admits continuous values for
eigenvalue of the Casimir operator and we shall give heuristic physical
argument below to support this claim. In other words, although the unitary
infinite dimensional representations of $SU(1,1)$ admit both
discrete and  continuous values for $j$ (and hence the Casimir), for the case
at hand we only need to pick up the solutions corresponding to the continuous
one.
 For convenience,  
introduce the continuous eigenvalue $\rho$ which appears in eigenvalue
equation as follows
\be
{\hat C}Y^m_{\rho}=(\rho ^2+{1\over 4})Y^m_{\rho}, ~~J_3Y^m_{\rho}=mY^m_{\rho}
\ee
There is an intuitive way to see why the continuous eigenvalue solutions are
the correct ones. For notational simplicity, momentarily, let us write
(\ref{lb}) as 
\be 
\label{continue}
O_{LB} \psi=\epsilon \psi
\ee
$O_{LB}$ is the Laplace-Beltrami operator defined in (\ref{lb}) and $\epsilon =
{\rho}^2 +{1\over 4}$ is the eigenvalue. Note that $\rho$ is responsible for
exponential behaviour of the wave function at infinity. The disc boundary
corresponds to $\alpha \rightarrow \infty$. If we want to see how $\psi$
behaves on the boundary, then (\ref{continue}) could be well approximated 
(note: ${\rm sinh }\alpha \sim {{e^{\alpha}}\over 2}$) by
\be
-\left({{{\partial}^2}\over {\partial {\alpha}^2}} +{{\partial}\over {\partial
\alpha}}\right)\psi=\epsilon \psi
\ee
Therefore, near the disc boundary 
\be
\psi \sim e^{-\alpha /2}e^{\pm i\rho\alpha}
g(\alpha)
\ee
 We have displayed the exponential $\alpha$-dependence in two terms explicitly
and $g(\alpha)$ is some smooth function. Also note that the probability
density in our curved space is $|\psi|^2{\rm sinh}\alpha d\alpha d\beta$
and this explains the first exponential. When $\epsilon \ge {1\over 4},~
\rho$ is real and second exponential is like a plane wave part. Let us assume
that we admit eigenvalues $\epsilon < {1\over 4}$, in that case $\rho$
becomes imaginary. Consequently, the wave function will grow exponentially in 
some direction. If it falls off exponentially in all directions, we may 
conclude that this solution corresponds to a bound state in a homogeneous space
where no interaction potential is present. On the other hand, 
 the exponentially growing solutions, alluded to earlier (
still $\epsilon <{1\over 4}
$ is being discussed), are not admissible as wave functions of any 
continuous spectra. We are led to conclude, from these reasonings, that
the eigenvalues of $O_{LB}$ on the pseudosphere, as being studied here, are
continuous and the lowest eigenvalue, $\epsilon _0 =+{1\over 4}$; and  
furthermore, the eigenvalues extend up to $\infty$.\\
Now we return to discuss the eigenfunctions of ${\hat C}$ and present explicit
solutions of the wave function.
We choose $m$ to take integer values. Recall that  the form of the differential
operator $\hat C$ is given by (\ref{lb}) and the compact operator $J_3=-i
{{\partial}\over {\partial \beta}}$, then we can write
\be
Y^m_{\rho}=e^{im\beta}X^m_{\rho} \ee
Here of course $\beta$ takes values $0~{\rm to}~2\pi$. The differential
equation satisfied by $X^m_{\rho}$ is the familiar one
\be
-{1\over {{\rm sinh}\alpha}}{\partial \over {\partial \alpha}}
({{\rm sinh}\alpha}){{\partial}\over {\partial \alpha}}X^m_{\rho} +{{m^2}
\over {{\rm sinh}^2\alpha}}X^m_{\rho}=(\rho +{1\over 4})X^m_{\rho}
\ee
The solutions are associated Legendre functions \cite{baniz,dfj} 
$P^m_{-{1\over 2}+i\rho}$. Thus, the normalized eigenfunction is
\be
\label{ylm}
Y^m _{\rho} (\alpha ,\beta)={{e^{im\beta}}\over {{\sqrt{2\pi}}}}
{{\Gamma (i\rho+{1\over 2}-m)}
\over {\Gamma (i\rho)}} P^m_{-{1\over 2}+i\rho}({\rm cosh} \alpha) \ee
Notice that the wave functions thus obtained belong to the continuous
series of the representations of $SU(1,1)$. We present below the
representation of the associated Legendre functions in terms of hypergeometric
functions
\be
P^m_{l}(w)={1\over {\Gamma (1-m)}}({{w+1}\over {w-1}})^{m\over 2}~
{_2F_1}(-l ,l +1;1-m;{{1-w}\over 2}) \ee
where $l =-{1\over 2}+i\rho$. Moreover, the canonical function
$P^m_{-{1\over 2}+\rho}(w)$, $m\in {\bf Z}$, have special properties
\be
P^m_{-{1\over 2}+i\rho}(w)=P^m_{-{1\over 2}-i\rho}(w)
\ee
which, we may recall, follows from the relation between $P^m_{l}$
and $P^m_{- l -1}$. Since $j=-{1\over 2}+il , ~l>0$ for the continuous
series, one may utilize this property the establish the orthogonality
relation
\be
|{{\Gamma ({1\over 2}+i\rho -m)}\over {\Gamma (i\rho)}}|^2 \int _1^{infty}
P^m_{-{1\over 2}+i\rho}(w)P^m_{-{1\over 2}-i\rho}(w) dw=\delta (\rho -\rho ')
\ee
The orthogonality relation is defined for $\rho , \rho '\in {\bf R^{+}}$
and $m\in  {\bf R}$. The completeness relation reads as
\be
\int _0^{\infty} |{{\Gamma ({1\over 2}+i\rho -m)}\over {\Gamma (i\rho)}}|^2
P^m_{-{1\over 2}+i\rho}(w)P^m_{-{1\over 2}-i\rho}(w')d\rho =\delta (w-w')
\ee
These relations are known as Mehler transformations \cite{baniz}.
Now, using the value of the Casimir in (\ref{wdw}), we arrive at the 
differential equation
\bea
\label{scalef}
\left({{d^2}\over {d a^2}}+{{d}\over {da}} -a^2 +{{\rho ^2+{1\over 4}
}\over {a^2}}\right)
{\cal U}(a)=0
\eea
The solutions are Bessel functions \cite{bessel}:
${\cal U}(a)_{\nu}=J_{\pm {i \nu}/2}({i\over 2}a^2) $
where ${ \nu}^2= \rho ^2+{1\over 4}$ is introduced for 
notational conveniences.Thus,
the solution,   $\Psi$,  is given by
\bea
\label{psi}
\Psi (a,\alpha ,\beta)={\cal U}(a)_{\nu}e^{im\beta}
P^m_{-{1\over 2} +i\rho }({\rm cosh}\alpha)
\eea

\noindent Let us discuss the necessary inputs for determining the wave function
from the available class of solutions. The $a$ dependent part of the
wave function admits two possibilities, 
in $a\ge 0$ region, where the wave function
corresponds to classical configurations such that either the Universe
is expanding or it is contracting. We should choose the solution which is
identified with an expanding Universe. Moreover, the wave function is
expected to be well behaves for large $a$. The first criterion (
expanding Universe i.e. Hubble parameter $H>0$ ) implies that in the limit
$a\rightarrow 0$ the wave function should have negative eigenvalue of the
momentum ${\hat P}_a$, since classically $H=-{{P_a}\over{ {a^2}}}$. The function
that fulfills this requirement is $H^{(1)}_{{i\over 2}\nu}(z)$, where 
$z={i\over 2}a^2= {{e^{{i\pi}\over 2}}\over 2}a^2 $. In the limit $a\rightarrow
0$ 
\be
H^{(1)}_{{i\over 2}\nu} \rightarrow {1\over {\pi}}\Gamma ({i\over 2}\nu)
e^{\pi \over 4}e^{-i\gamma} \ee
where $\gamma =ln a$ introduced for brevity of notation.
 Note that $\dot {\gamma} = H
=-{{P_a}\over {a^2}}$ and $P_{\gamma}=-{3\over 2}e^{3\gamma}{\dot \gamma}$
and therefore, $H>0$ implies that wave function should have negative
eigenvalue of ${\hat P}_{\gamma}$. Indeed, if we identify $P_{\gamma}=-i
{{\partial}\over {\partial \gamma}}$, then the Henkel function in the limit
of $a\rightarrow 0$ gives the requisite eigenvalue.
For large values of $a$ i.e. $a\rightarrow \infty$, the behaviour is
\be
H^{(1)}_{{i\over 2}\nu}\rightarrow~{\rm Const} {{e^{-{{a^2}\over 2}}}\over a} 
\ee
which is well behaved for asymptotic values of the scale factor.\\
The wave function (\ref{psi}) is in the factorized form, where the 
dependence on the scale factor $a$ given by the Hankel function, 
$H^{(1)}_{{i\over 2}\nu}({i\over 2}a^2)$ and the dependence on $\chi$ and
$\phi$ (through $\alpha$ and $\beta$) are given by the associated
Legendre polynomial $P^m_{-{1\over 2}+i\rho}({\rm cosh}\alpha) $
and the phase factor $e^{im\beta}$. Notice that $\alpha$ and
$\beta$ are related to $\chi$ and $\phi$ through equation (\ref{axdilrl}).
It is not very easy to extract the behaviour of the wave function in the
strong or in the weak coupling limit although the behaviour of the
associated Legendre polynomial is well known for small or large values of
its argument ${\rm cosh}\alpha$. Therefore, it is useful to look at the
solution to the axion-dilaton part of the wave function in terms of the
set of variables $\lambda _1$ and $\lambda _2$ introduced in the previous
section.\\
We may recall the introduction of the complex field $\lambda=\lambda _1 +
i\lambda _2$. We write
\be
\lambda =\lambda _1+i\lambda _2 \equiv  x+iy
\ee
and therefore, $x=\chi$ and $y=e^{-\phi}$. The Hamiltonian, $\tilde H$, satisfies
the constraint

\bea
\label{hxy}
{\tilde H}={1\over 2} ({{{P_a}^2}\over a} +a -{{ y^2}\over {a^3}}(P_x^2+P_y^2))=0
\eea
This is the classical Hamiltonian constraint. $P_x$ and $P_y$ refer to the
canonical conjugate momenta of the coordinates (moduli) 
$x$ and $y$ respectively.
In passage to WDW equation, we represent  $P_a, P_x$ and $P_y$ as differential
operators and also take into account operator ordering as alluded to above.
The wave function can be expressed in the factorized form.\\
Let us focus on the part of the Hamiltonian depending on $x$ and $y$ coordinates
\be
{\tilde H}(x,y)={1\over 2}y^2(P_x^2+P_y^2)
\ee
We immediately recognize that this Hamiltonian
descibes motion in the Poincar\'e upper half
plane since $-\infty \le x \le +\infty$ and $y=e^{-\phi}$ takes only positive
values. The eigenvalue equation for this Hamiltonian has been studied in detail
from several perspectives \cite{gutz}. Let us denote the wave function
as $\psi ^{UP}(x,y)$ and it satisfies the eigenvalue equation
\be
\label{peigen}
{\tilde H}(x,y)\psi ^{UP}(x,y)=E_P\psi ^{UP}(x,y)
\ee
where $E_P$ is the eigenvalue for the Poincar\'e Hamiltonian. We recall that
the problem at hand is analogous to motion of a particle on surface of constant
negative curvature as was pointed out when we formulated the problems in terms
of the coordinates $v_i$. We are examining the same problem now in variables
$x$ and $y$. \\

The eigenfunction and the eigenvalues are given by
\bea
\label{wfnp}
\psi ^{UP}(x,y)={\sqrt {{\rho {\rm sinh}\pi\rho}\over {{\pi}^3}}}
{\sqrt y} K_{i\rho}(|\kappa |y)e^{i\kappa x}
\eea
 and
\be
E_{P}=({\rho}^2+{1\over 4})
\ee
Here, $\rho >0$ and zero eigenvalue for $\kappa$ is excluded i.e.
$\kappa \ne 0$.
The wave function, $\psi ^{UP}(x,y)$, satisfies the orthonormality condition
 \bea
\int _{-\infty}^{+\infty} dx\int _0^{+\infty} {{dy}\over {y^2}}
\psi ^{UP}_{\rho ,\kappa}(x,y)\psi ^{{UP}\star}_{\rho ',\kappa '}(x,y)
=\delta (\rho -\rho ')\delta (\kappa - \kappa ')
\eea
and the completeness relation is
\bea
\int _{-\infty}^{+\infty}d\kappa \int _0^{+\infty}d\rho \psi ^{UP}_{\rho ,\kappa}
(x',y')\psi ^{{UP}\star}_{\rho ,\kappa}(x,y)=yy'\delta (x-x')\delta (y-y')
\eea
Note the appearance of $yy'$ which is due to definition of the line element
in the upper half Poincar\'e plane.\\
Let us discuss some of the features of the wave function $\psi ^{UP}$. We 
recall that the Hamiltonian $\tilde H$ contains only $P_x$ and has no explicit
$x$-dependence, whereas it depends on $P_y$ as well as on $y$. Thus, we expect
that the wave function will have a plane wave component in $x$ 
(which amounts to a plane wave in the axion field) to due the 
translational invariance in this coordinate. 
We remind the reader that the action
(\ref{axd}) depends on axion field through its derivative and consequently, 
the equations of motion for $\chi$ is a conservation law. This is manifest in
our solution. The $y$-dependent part of the wave function contains the
Bessel function $K_{i\rho}(|\kappa |y)$, besides the $\sqrt y$ factor.
Next, we look at the behaviour of the wave function in the limit
$\phi \rightarrow \infty$ which corresponds to the strong coupling limit and
this amounts to taking $y\rightarrow 0$ limit. The Bessel function in this
limit has the behaviour ${\rm lim}_{y\rightarrow 0} ~~K_{\nu}(y) \sim {1\over 2}
\Gamma (\nu) ({z\over 2})^{-\nu}$. We remind the reader that $y=e^{-\phi}$ and
$\nu =i\rho$ in our case. Thus, the Bessel function looks 
like a plane
wave in this limit. Therefore, the wave function, $\psi ^{UP}(x,y)$,
 behaves like 
a plane wave in both axion and dilaton
with an additional  factor of $\sqrt y$.\\
Now we  consider the solution to (\ref{peigen}) in a special case where
the wave function $\psi _{UP}$ is $x$-independent. We would like to
examine how the wave function is related to solutions in the semiclassical
approximation in this simple example.
Notice that 
 case the wave equation is the Helmholtz equation $-y^2{{d^2\psi}\over
{dy^2}}=\epsilon \psi$ and the two solutions are $y^{{1\over 2}+i\rho} ~
{\rm and } ~y^{{1\over 2}-i\rho}$ and $\epsilon = \rho ^2 +{1\over 4}$, as 
before. For $\epsilon >{1\over 4}$, the solution 
\be
\label{onlyy}
y^{{1\over 2}+i\rho }
=y^{1/2}e^{i\rho {\rm ln}y}
\ee
We remark, in passing, that $x$-dependent solutions may be generated from
the wave function (\ref{onlyy}) by implementing a suitable S-duality  
transformation given by
\bea
\label{pslr}  
  z \rightarrow z'={{az+b}\over {cz+d}},~~{\rm with}~ad-bc=1
\eea 
and the imaginary part of $z'$, $y'$, is given by
\be
y'={{y}\over {|cz+d|^2}}
\ee
Thus the transformed wave function will carry $x$-dependence.
   The following interpretation of the wave function
is worth noting as we go over to the classical limit. 
The $\hbar \rightarrow 0$ limit
corresponds to  $|\rho |\rightarrow \infty$. This can be seen as follows:
if we restore the all dimensionful parameters such as radius of the
pseudosphere, $\cal R$, Planck's constant and the mass of the particle $ m_p$,
 then we can define an energy scale (in quantum mechanics) 
$e_s={{\hbar}\over {
2{ m_p} {\cal R}^2}}$. Thus energy can be expressed as $E=e_s \zeta$ in terms
a dimensionless 'eigenvalue',$\zeta$. Therefore,
 the semiclassical limit is equivalent to
sending $\zeta $ to $\infty$. This is also same as sending $\rho$ to asymptotic
values. From semiclassical arguments (say like WKB)
\be
\psi \sim Ae^{iS}
\ee
We can identify ${\rm ln}~ y$ as the classical action since it is $\int _1^y ds
$ along the $y$ axis for the case at hand i.e. $x=0$ is chosen. Moreover, 
$\sqrt y$ is the invariant density in the Poincar\'e metric. Therefore, we
can identify $\psi (y)$ as the solution in the WKB approximation.
\\
Let us write down the full wave function as a solution to WDW equation
in terms of the scale factor, $a$ and the Poincar\'e coordinates $x$ and $y$:
\be
\Psi (a,x,y) =H^{(1)}_{{i\over 2}\nu}({i\over 2}a^2)\psi ^{UP}(x,y)
\ee 
This wave function is well behaved in the strong coupling limit. 
If we look at the classical solution
to axion-dilaton cosmological solution, the dilaton approaches strong coupling
regime as $t$ approaches zero and in that limit the scale factor tends to zero.
We should keep in mind, however, that  the tree level string effective is 
not quite
reliable in this domain and it might be necessary to add higher order terms
in curvatures and additional stringy correction terms to the tree level
effective action. One possible way to deal with solution at the strong coupling
domain is to apply an S-duality transformation so that we go over to the
weak coupling regime. This might be good strategy to adopt since we are
dealing with S-duality invariant action.  It is important to note
 that we have succeeded in 
presenting an exact and complete 
solution to WDW equation for axion-dilaton string cosmology
which was lacking so far.

\section{ Hidden Symmetry in Axion-Dilaton Cosmology }

We have  argued that the axion-dilaton string cosmology might
be endowed with  higher symmetry  since the string effective actions
in lower dimensions exhibit enhanced symmetries. 
The string effective action depends only on the cosmic time coordinate in
the cosmological models.
 In our previous  note \cite{j1}, we
presented evidence for existence of a $w_{\infty}$ algebra
 in the form proposed  by Bakas \cite{bakas}. Our result was
based for  a specific choice of the Casimir, $C=-{3\over {16}}$ of the
S-duality group $SU(1,1)$.  It was argued that unitary representations of
$SU(1,1)$ are infinite dimensional. Thus the wave function associated with
the WDW equation for axion, dilaton and graviton  is
highly  degenerate.  Moreover, 
the generators of $SU(1,1)$ commute with the 
action and hence with  the Hamiltonian due to 
S-duality invariance of the theory.
For the specific choice of the Casimir, $(C= -{3\over {16}})$, we expressed
 the raising, lowering generators $J_{\pm}$ and the compact diagonal
generator $J_3$ in terms of the creation and annihilation operator of a single
boson. We recall that
\bea
\label{oneb}
J_{+}={1\over 2}(a^{\dagger})^2 ,
 ~J_-={1\over 2}a^2 ,
~ J_3={1\over 2}(a^{\dagger}a
+1/2)
\eea
$[a, a^{\dagger }]=1$. Defining,
$|n>=(n!)^{-1/2}(a^\dagger)^n|0>$, where $a|0>=0$ is the condition on vacuum,
we find
\bea
\label{raise}
J_+|n>={{\sqrt {(n+1)(n+2)}}\over 2}|n+2>,
\eea
\bea
J_-|n>={{\sqrt {(n(n-1)}}\over 2}|n-2>,
\eea
and
\bea
 J_3|n>=(n+{1\over 2})|n>
\eea                                                                             
This representation for $J_{\pm}$ and $J_3$ is due to Holman, Biedenharn and
Ui \cite{ui}. In this case, we get  two
different representations of the Lie algebra
from  $|n>$: (i) For odd $n$,  $j=-3/4$
and (ii) for even $n$, $j=-1/4$ which belongs to the supplementary series.
When $J_3$ and $J_{\pm}$ are defined
as in (\ref{oneb}),
a  suitable set of operators can be constructed, in both the cases,
which satisfy the $w_{\infty}$ algebra \cite{bakas}
\bea
\label{w}
[{\cal V}_m^{(k)} ,{\cal V}_n^{(l)}] &=& \left((k+1)(n+1)-(m+1)(l+1)\right)
{\cal V}
_{m+n}^{(k+l)}
\eea
where ${\cal V}_{m}^{(k)}=(a^{\dagger})^{m+1}a^{k+1},~n,m \ge -1$. Here,
 we have computed  the classical algebra, ignoring
normal orderings.\\
The presence of a $w_{\infty}$ in axion-dilaton cosmology was unveiled in
a heuristic manner for a special choice of the Casimir so that the 
generators of $SU(1,1)$
 have a very simple representation in terms of a single boson operator.
In recent years $w_{\infty}$ and $W_{\infty}$, which might
be considered as deformation of the former,  have played useful roles in
a variety of problems
in physics. They first appeared  as $N\rightarrow \infty$ limit
of $W_N$ algebras \cite{wn}. These algebras were studied in the context of 
$c=1$ theories \cite{c1}, in large N limit of $SU(N)$gauge theories \cite{g1}
 and in quantum Hall effect \cite{qh}. 
In what follows, we shall summarize some of the essential aspects of W-infinity
algebras and then argue that axion-dilaton cosmology also shares
those features. These algebras may be conceived as $N\rightarrow \infty$ limit
of $W_N$ algebras. The $W_N$ algebras are extended Virasoro algebras with a
field content of higher conformal spins: $2,3,...N$. There are nonlinear terms
in the commutation relations and such terms get more and more complicated as
N takes higher and higher values. Therefore, these algebras do not correspond 
to Lie algebras so long as N is finite. When the limit $N\rightarrow \infty$
is taken, it is argued that the algebra assumes a rather simple form 
\cite{bakas}. Of
course, the passage to $N\rightarrow \infty$ limit is not unique; and
consequently, different algebras emerge depending on how one first rescales
the generators and the structure constants of $W_N$ algebras. Bakas proposed
a prescription for the $N\rightarrow \infty$ limit to arrive at an algebra,
known as $w_{\infty}$ 
\bea
\left[w_m^{(i)} , w_n^{(j)} \right]=\left[(j-1)m-(i-1)n \right]
w_{m+n}^{(i+j-2)} 
\eea
Note that $w_m^{(i)}$ are generators of conformal spin $i$. In this case
the algebra has central charge {\it only} for $i=2$ which turns out to be the
Virasoro algebra. On the other hand, in the case of $W_N$ algebra, there is
central term for each of the conformal spin. 
Pope, Romans and Shen \cite{cp}proposed
the $W_{\infty}$ algebra which may be interpreted as the deformed $w_{\infty}$
 algebra which has central term for all conformal spins. The Jacobi identity 
is imposed to derive the structure constants. They
show that $W_{\infty}$ algebra has a subalgebra whose generators are conformal
spins $2,3,4,...$ in the $\underline 3$, $\underline 5$, $\underline 7$...
representations of $SL(2,R)$. The resulting algebras are anomaly free. One
constructs tensor operators of $SL(2,R)$, ${\cal T} (C)$,  where
 $C$ is the value
of the Casimir generator of $SL(2,R)$. In order to study the subalgebras
of the $W_{\infty}$ algebra, called wedge algebras $W_{\wedge}$, the 
generators $V_m^{(i)}, |m|\le i+1$ are identified.  The commutators of the
generators belonging to this subset gives another generator which also belongs
to the same subset. The $W_{\wedge}$ is anomaly free. The generators belonging 
to $W_{\wedge}$ are constructed as family of tensor operators which are
in $SL(2,R)$ satisfying special properties. The starting point is to consider  
the three  generators of $SL(2,R)$, 
$\{L_{\pm} , L_0 \}$, which transform as $\underline 3$. We 
consider higher tensor operators $T_m^l,~ -l\le m \le l$ transforming 
as the $(2l+1)$-dimensional representations of $SL(2,R)$. Such tensors are
constructed from the polynomials of degree $l$ in the generators $L_{\pm}$ and
$L_0$. One of the prescription is to start from the highest weight state
$T_l^l \equiv (L_+)^l$. We act with the step down operator $L_-$ successively
in order to construct lower weight states. After $(l-m)$ such operations,
as specified below, one arrives at
\bea
\label{ttensor}
T_m^l={1\over{(-2l)_{l-m}}}\left(Ad_{L_-}\right)^{l-m}(L_+)^l
\eea  
where $Ad_X(Y) =[X,Y]$; therefore (\ref{ttensor}) amounts to taking $(l-m)$
commutators of $L_-$ with $(L_+)^l$ i.e. $[L_-,[L_-,[L_-...(L_+)^l]]]]]$.
The denominator $(-2l)_{l-m}$ has the meaning $(a)_n={{(a+n-1)!}\over {
(a-1)!}}$.
We remind the reader that $[L_{\pm},L_0]=\pm L_{\pm}$ and $[L_+ ,L_-]=2L_0$.
Therefore, we can expand $T_{\pm m}^l$ as polynomials in $L_0$ times $(L_{\pm})^m$
which have coefficient expressed in terms of the Casimir of $SL(2,R)$.
\bea
T_{\pm m}^l=\left(L_0^{l-m}\pm {1\over 2}(l-m)L_0^{l-m-1}+...\right)L_{\pm}^m
\eea
The normalization is so adjusted that the first term has coefficient unity.
As is the case with $SU(1,1)$, $SL(2,R)$ has infinite dimensional unitary
representations. For a given choice of the value of the Casimir, say C, 
 $T_m^l$, the set of tensor operators  constructed above, close into an 
infinite dimensional algebra labeled as ${\cal T}(C)$. It is argued that the
wedge algebra coincides with ${\cal T}(C)$ for some appropriate value of $C$.
One finds that the wedge algebra contained in $W_{\infty}$ is $SL(2,R)$
operator algebra ${\cal T}(C=0)$; the Casimir takes zero value. Arguments for
the connection between wedge algebra and the $SL(2,R)$ operator algebra
due to Pope, Romans and Shen \cite{cp} are based on abstract constructions.
Since $SL(2,R) \sim SU(1,1)$, the above argument will go through for our
cosmological problem. In fact, we could have adopted $SL(2,R)$ as our S-duality
group as well. Notice that we have realizations of the generators (of 
$SU(1,1)$)
\bea
\label{jop}
J_{\pm}={\mp}e^{\pm i\beta}{{\partial }\over {\partial \alpha}} -i~{\rm coth}
\alpha ~e^{\pm i\beta}{{\partial }\over {\partial \beta}},~~~J_3=-i{{\partial}
\over {\partial \beta}}
\eea                                                                            in terms of the coordinates of the pseudosphere (in turn the axion and 
the dilaton). Therefore, the set of operators $T_m^l$ can be constructed
from the generators (\ref{jop}). We recall that our eigenfunctions 
correspond to the continuous series representation of $SU(1,1)$. Consequently,
when we fix the value of the Casimir, the magnetic quantum number takes
values, $m=0, \pm 1, \pm 2, \pm 3,....$ as mentioned earlier. We can construct
the operators $T_{\pm m}^l$ according to our convenience and the algebra
will be realized on the infinitely degenerate eigenstates.  
Furthermore, the exact eigenfunctions 
have also been constructed by us in
this case. We are led to the conclusion that in this case we have a physical
realization of the algebra proposed by Pope, Romans and Shen \cite{cp}.
\\
We would like to discuss a limit when the Casimir, $C\rightarrow \infty$.
This corresponds to the large eigenvalue of the quantum mechanical equation
on the pseudosphere. This limit is of relevance due to the fact that the
semiclassical approximation can be taken in the large C regime. In study
of quantum mechanics on the pseudosphere, the inverse of the eigenvalue
corresponds to ${\hbar}^2$ and $\hbar \rightarrow 0$ (the semiclassical limit)
amounts to studying the system for large eigenvalues. It is important in
the context of the study the chaotic properties of 
the motion on the pseudospehere as
we shall briefly discuss in the following section. Furthermore, the algebra
of the operators in large C limit also takes rather simple form.
If we define the set $\{{\hat{T}}_m^{(i)} \}$ to be the suitably 
rescaled generators from
the set $\{T_m^{(i)} \}$, then the algebra becomes
\bea
\label{cinf}
\left[ {\hat T}_m^{(i)} , {\hat T}_n^{(j)} \right] =
\sum _k{{{{\sqrt {2k+1}}(i+j+k+2)
{\Delta}^2({i\over 2},{j\over 2},{k\over 2})}\over{\Delta (i,j,k)}}} C^{ijk}
_{m,n,m+n}{\hat T}_{m+n}^{(k)}
\eea
where
\be
{\Delta}(a,b,c)\equiv {\sqrt{{(a+b-c)!(b+c-a)!(c+a-b)!}\over{(a+b+c+1)!}}}
\ee
and $C_{mnp}^{ijk}$ are Clebsch-Gordan coefficients whose properties can be
found in ref. \cite{bdlk}. In (\ref{cinf}), the sum over $k$ is over all 
values of k with the constraint that $i+j+k$  odd and the Clebsch-Gordan
coefficient $C^{ijk}_{m,n,m+n}$ have nonzero values. Notice the absence any
central term in the algebra (\ref{cinf}). As we have argued, large 
eigenvalue limit is the semiclassical limit and therefore, central terms, if
any, are not expected to make their appearances in these algebras.
\\
It is worth while to mention that  the novel symmetry,  associated with 
cosmological scenario of axion, dilaton and graviton, owes its origin to
the existence of the S-duality group $SU(1,1) \sim SL(2,R)$. The highly
degenerate wave function in the axion-dilaton sector exists due to the
same reason.

\section{Axion-Dilaton Cosmology and Chaos }

In this short section, we briefly discuss
chaos in axion-dilaton cosmology. This is another interesting aspect.
The axion-dilaton moduli parametrize a space of constant negative curvature as
is evident from its parametrization as the surface of a pseudosphere or its
parametrization as Poincar\'e upper half plane. Motion on such a surface is known to be chaotic. However, in the cosmological scenario the moduli are coupled
to gravity; in our case it is the FRW metric. \\
The study of chaos, in the context of cosmology, 
started  more than three decades ago \cite{bkl,mis} and an early review
is contained in the article by Barrow \cite{bar}. Misner \cite{misner} and
Chitre \cite{chitre} studied  chaos in 
Bianchi type VIII and IX cosmologies. When one is
sufficiently close to the singularity, in a suitable coordinate system,the
problem can be reduced to a special two dimensional billiard on a homogeneous
space of constant negative curvature. Recently, the billiard problem for
Einstein-dilaton gravity along with p-form fields has been studied in detail
and we refer the reader to the article of Damour, Henneaux and Nicolai 
\cite{dhn}.\\
Let us turn our attention to string theory.
In string theory, we encounter moduli spaces where the generic 
characteristics is such that 
 a noncompact group is modded out by its maximal compact
subgroup and eventually realized as a discrete subgroup. The well known
example is compactification of a bosonic string \cite{jjhs}
 ( we illustrate it as a simple case)
on a d-dimensional torus, $T^d$. The moduli, $\cal M$ parametrize ${O(d,d, Z)}
\over {O(d)\otimes O(d)}$. Thus the effective action can be expressed as 
nonlinear $\sigma$-model in the target space and
is invariant under global $O(d,d,Z)$ and local $O(d)
\otimes O(d)$. From the point of view of the evolution of the string,
we usually encounter
  situations when a discrete group $\Gamma$ connects
different configurations of the spacetime background field in the action and 
then  the sigma models correspond to same
conformal field theory. 
The S-duality group is ${SL(2,R)}\over {U(1)}$ (we remind
that $SL(2,R) \sim SU(1,1)$) and eventually $SL(2,R)$ is reduced to $SL(2,Z)$
in full quantum theory. Moreover, as mentioned in the introduction, vacuum
expectation values of the dilaton and the axion are related to the 
coupling constants and the VEV of the axion
is related to the $\theta$-angle that appears as coefficient of the topological
term of gauge theories in the four dimensional action.\\
The solution to equations of motion in Poincar\'e half plane is well 
understood. The evolution of $\lambda \equiv x+iy$ corresponds to  semicircular
trajectories lying in the upper half Poincar\'e plane. The center of the
semicircles lie on the real axis; the location of the center and radius of
the semicircles are determined from the initial conditions supplied to
solve the classical  equations of motion. In the cosmological
scenario, as discussed in section 2, we aim at solving coupled set of equations
involving metric and the matter fields. 
When we consider axion-dilaton evolution in the cosmological context (
with FRW metric), the equation of motion (\ref{friction}) is modified due the
presence of the Hubble parameter, H, which couples to $\dot {\xi}$. Thus
it is like a frictional force - this is already well known from the evolution
equation of the scalar field, for example in models of inflation. 
\\
Let us briefly recapitulate the situation for the classical dynamics and
quantum equations of motion in our problem ( motion on the pseudosphere):
(i) classical equations are coupled differential equations. (ii) The WDW 
equation is solved by adopting the method of separation of variables and
we have a quantum mechanical eigenvalue equation for axion-dilaton. Horne
and Moore \cite{hm} persuasively argue that classical chaotic behaviour
continues to be exhibited by axion-dilaton in the presence of gravity
- when the frictional force of Hubble parameter is present. 
Usually, the classical chaotic
motion is  characterized by two attributes: (a) The phase space trajectories
diverge exponentially and (b) the admissible phase space volume be compact. 
It is important to note that when the motion is unrestricted on the
 pseudosphere, the trajectories still diverge. In order to satisfy the
periodicity properties of the trajectories/orbit one has to construct the 
compact
surfaces the through introduction of the notion of
suitable periodic boundary conditions in this curved space; just as we
construct a torus by identifying opposite sides of a rectangle in a two
dimensional Euclidean plane.
This is the way to get finite volume phase space.\\
In the quantum mechanical considerations, we do not discuss  particle
trajectories/orbits since we solve for wave functions. The connection with
chaotic behaviour may be established in the semiclassical regime. In view
of the above discussions, we may argue that semiclassical solutions to the wave
equation on unrestricted pseudosphere might be able to bring out one of
the features of chaos. When we solved for the wave equation in terms of
 variables $\alpha$ and $\beta$, we chose the wave function to be periodic in
$\beta$ since the compact generator $J_3$ was required to be diagonal. The
large eigenvalue limit is the classical limit where $\rho \rightarrow$ takes
asymptotic values. Consequently, the wave function, associated Legendre 
polynomial is to be evaluated in large $\rho$ limit. 
\\
Now, we  proceed to discuss  solutions to wave equation in the semiclassical
 approximation and its connection with the chaotic behaviour of classical
theory. We have shown in section 3 that plane wave 
solutions are ($E_U \sim ~{\rho}^2$ for large $\rho$)
\be
\label{cwkb}
\psi _U \sim {\sqrt y} e^{\pm i{\sqrt{E_U}}{\rm log}y}
\ee
This solution is expressed in terms of the coordinates of Poincar\'e upper  
half plane. These plane waves are WKB solutions \cite{balv,gutz}.
Therefore, we may identify the action $S={\pm} {\sqrt E_U}{\rm log}~y$ and
$A^2(S)=y=e^{-{S\over{\sqrt{E_U}}}}$; here S being the action that appears
in the exponential when we derive the WKB wave function and $A(S)$ is the
coefficient of the exponential. If we define $\tau ={S\over {2E_U}}$, 
then, $A^2(\tau )=e^{-{\sqrt {2E_U}}\tau}$. 
$A^{-2}(S)$ measures divergence of trajectories in the semiclassical approach 
to understand chaos in the surface of constant negative curvature (Poincar\'e
upper half plane) \cite{balv}. Thus one extracts the Liapunov exponent by
examining large S behaviour of $A^{-2} (S)$ 
In our convention, $E_U$ carries
dimension of inverse time and is related to the Liapunov exponent $\omega$.
If we are to compare with the convention followed in literature for the study
of chaos \cite{balv,gutz} for the exponent $\omega$, we should set $E_U=2$ which
gives $\omega =2$. This is the value obtained for Liapunov exponent
for exponential divergence of classical trajectories in the study of chaos
in Poincar\'e upper half plane \cite{balv,gutz}. Thus the WKB approximation
reproduces one of the features of chaos in the classical dynamics. We 
would like to remind the reader that the correspondence is for chaos of
$x,y$ coordinates when motion is in flat Minkowski background. However, it
is not obvious, although axion-dilaton cosmology is argued to be chaotic
\cite{hm}, that the classical trajectories will diverge with the same exponent
as the flat spacetime case. Nevertheless, it is quite encouraging that the
quantum cosmological solutions, in the semiclassical approximation, does
bring out one of the features of chaos.\\
We end this section with following comments. In order to study chaos on the 
pseudosphere, we have to introduce periodic boundary conditions. This
is achieved following a set of prescriptions as discussed below. An immediate
consequence of the periodic boundary condition on the compact surface,
constructed from pseudosphere, for the quantum theory is that the energy
eigenvalue becomes discrete. Moreover, for the unconstrained pseudosphere,
the energy eigenvalue and wave functions can be solved exactly. The 
compact surface with negative constant curvature, obtained from the 
pseudosphere, corresponds to a Riemann surface of at least genus two, $g=2$,
although higher genus Riemann surfaces can give rise to constant negative
curvature surfaces. It is a formidable task to solve for eigenfunction and
eigenvalues of the Hamiltonian which satisfy `
periodic' boundary conditions \cite{balv,gutz}.
There are no exact straight forward solutions to the general problem.
The problem is further complicated in the cosmological context since one has
to account for the presence of gravity while addressing the question of
chaotic phenomena. We feel that considerable amount of efforts are necessary
to provide a complete understanding of chaos in axion-dilaton cosmology.

\section{Summary and Discussions }

We have investigated various aspects of  axion-dilaton string cosmology 
in this article. As a first step, before addressing the cosmological problem,
we studied the S-duality properties of string effective action in four
spacetime dimensions and constructed (classical) N\"other currents and
corresponding charges which are the generator of S-duality group. Axion and
dilaton parametrize the coset ${{SU(1,1)}\over {U(1)}}$. We considered
isotropic, homogeneous, FRW metric as our cosmological model. The moduli
depend only on the cosmic time coordinate, $t$. When one solves for the
classical equations of motion, there is curvature singularity, corresponding
to bing bang and the dilton goes to $\infty$ in this limit. Such a feature
of the solutions is observed for closed, open and flat Universe. It
was shown that axion-dilaton parametrize the surface of a pseudosphere
in a suitably defined coordinate system. Therefore, the action, in the
cosmological context for axion and dilaton, can be cast 
in a form which is analogous to
the Lagrangian of a particle moving on the surface of a pseudosphere in the
presence of FRW spacetime metric. \\
The Wheeler De Witt equation allows separation of the variables and
consequently, the wave function is factorisable as a product of two
functions; one depending on the scale factor and the other on the moduli. The
S-duality invariance of the action leads to the Hamiltonian which is sum of two
pieces; one which depends only on the Casimir of the S-duality group 
and a second part corresponds to  the Ricci scalar, computed in the
Einstein frame metric. Thus, we can solve for the 
wave function of the moduli from
the study of the representation theory of the S-duality group. The 
solutions correspond to the continuous representation of $S(1,1)$. 
 Therefore, the solution of the WDW equation, i.e.
${\hat{\cal H}} \Psi =0$ is infinitely degenerate. In other words,
the  zero eigenvalue solution to wave equation has infinite degeneracy. This
is a consequence of the fact that unitary representations of noncompact groups
are infinite dimensional.
 It is believed
that the exact symmetry is the discrete group $SU(1,1, Z) \sim SL(2,Z)$. We
have solved WDW equation without introducing a potential for the axion and
dilaton. A potential for the moduli could be generated due to nonperturbative 
effects 
 such as gluino condensation, in the cosmological
action. Horne and Moore  have  put forward proposals to construct
 such a potential on general
grounds \cite{hm}. They impose some  
constraints in deriving the form of the potential: that 
it respects discrete S-duality symmetry  and 
perturbation theory be valid in the weak coupling regime. It is
quite plausible that the gluino condensation occurs much below the regime of
quantum cosmology. In this respect, we may justify our solution of WDW
equation in absence of a potential for the moduli. Nevertheless, 
it is desirable
to look for a mechanism which lifts the degeneracy of the wave function
so that one obtains a unique ground state solution.\\
We have provided solutions to WDW equation in another set of coordinates where
the moduli define the upper half complex plane - the Poincar\'e coordinates.
This choice has two interesting features. First, behavior of the wave function
of the moduli, when dilaton takes positive asymptotic values can be studied
in a simple manner. This is the strong coupling limit of the theory. Moreover,
as we have argued earlier, when the scale factor
tends to zero (i.e. $a(t) \rightarrow 0$ with $t\rightarrow 0$) 
in the classical solution,  $\phi \rightarrow \infty$ in the same limit. Thus,
the behavior of the wave function near the curvature singularity 
(of classical solution) is rendered transparent. The second feature is that
the semiclassical approximation could be achieved efficiently since
the large eigenvalue limit is identified as the WKB limit and the corresponding
action can be read off immediately from the form of the wave function in
the Poincar\'e coordinates. \\
We have proposed \cite{j1} that there is an observed tendency that the 
lower dimensional string effective actions
are endowed with enhanced symmetries and the symmetry groups are enlarged
as we go to lower and lower dimensions. The cosmological axion-dilaton action
depends only on a single coordinate - the cosmic time. We  argued that there
is an underlying $w_{\infty}$ algebra for a specific choice of the Casimir
\cite{j1}.
In this article we provide a construction of $W_{\wedge}$ algebra 
 following the prescriptions of ref.\cite{cp},  It is shown that the algebra
assumes a simple form, without central terms, when the Casimir takes 
asymptotic values; which is the semiclassical limit as alluded to earlier.
We find the discovery  of W-infinity algebra in the context of axion-dilaton
 cosmology  to be one of the novel outcomes of our investigation.
 It is worth while to point out that the presence of
large exceptional groups and appearance of Kac-Moody algebras have been  
noticed in the context of cosmological billiards \cite{dhn} and this result
is another evidence in favour our conjecture that lower dimensional effective
actions possess higher symmetries.\\ 
The 5th section was devoted to study chaos in axion-dilaton cosmology. We 
showed, from qualitative arguments, that the system exhibits chaos in the
semiclassical limit. The semiclassical limit amounts to determining 
 wave functions
for large eigenvalues.  This is same as taking $\hbar \rightarrow 0$ limit.
We are encouraged to find the supportive evidence in favour of chaos from
our semiclassical analysis. Classical axion-dilaton cosmology
 is known to be chaotic as has been argued in ref.  \cite{hm}. We are aware 
that a rigorous study of chaos in quantum axion-dilaton cosmology will
require resolution of many difficult mathematical problems and the results
presented here are qualitative.
We feel that  our preliminary results and our attempts to study 
chaos in quantum string cosmology
will initiate further investigations in these directions.

\bigskip

\noindent {\bf Acknowledgments:} I have benefited from several of my colleagues
during the course of this work for their advice, suggestions and criticisms. I
would like to acknowledge valuable interactions with  Jan Ambjorn, Ed Copeland,
Avinash Dhar, Francis Ezawa,  Romesh Kaul, Ashoke Sen, Sandip Trivedi, Gabriele Veneziano
and Cosmas Zachos. I would like to thank Ashok Das,  Al Mamun and
Arsen Melikyan for helpful
communications. 

\newpage

\section{Appendix }

We summarize some of the properties of $SU(1,1)$ as useful information. This
is group of 2-dimensional, unimodular matrices and the Lie algebra of its
generators $\{J_1 ,J_2 , J_3 \}$ is given by (\ref{cr})
\be
[J_1 ,J_2]=-iJ_3 ,~~[J_2 ,J_3]=iJ_1 ,~~ [J_3 ,J_1]=iJ_2
\ee                                                                             and the Casimir is $-J_1^2-J_2^2+J_3^2$. We have already given the explicit
realization of the generators in terms of the Pauli matrices in section 2 and 
those generators generate nonunitary representations of $SU(1,1)$. In order to 
generate unitary representations of the group; we have to provide a realization
of the Lie algebra in terms of Hermitian operators. Thus, we should have an
 infinite dimensional vector space, because all irreducible representations
of noncompact Lie groups are infinite dimensional. Let us start with unitary
representation of $SU(1,1)$ in a Hilbert space ${\bf {\cal H}}$ and it will
be infinite dimensional. The matrix $S=\pmatrix{\alpha & \beta \cr {\beta}^{
\star} & {\alpha}^{\star} \cr} ~~\in SU(1,1)$. This has a maximal compact group
$U(\delta)=\pmatrix{e^{i{\delta \over 2 }} & 0 \cr 0 & e^{-i{\delta \over 2}} 
\cr }$. Now
${\bf {\cal H}}$ can be decomposed with respect to $U(\delta)$ into direct sum
of one dimensional representation $|m>$.
\be
{\bf{\cal H}}=\oplus|m>, ~~~U(\delta)|m>~=~e^{i m\delta}|m>,~~~|m>~
\in {\bf{\cal H
}}
\ee
$\delta =0$ mod $4\pi$ is the identity of the group; $m\in {\bf Z},~ {\rm or}
~ m\in {\bf {Z+{1\over 2}}}$. Note that quantization of $m$ does not follow
from the Lie algebra (unlike the compact case), but from the global properties.
 In decomposition ${\bf {\cal H}}=\oplus |m>$ each $m$ appears only once, since
the Casimir and $J_3$ form a complete system of commuting observables.\\
In section 3, we have briefly discussed the classification of the 
representations of $SU(1,1)$. The representations come in two types depending
on the choice of magnetic quantum numbers: (i) Integral magnetic quantum
number, and (ii) half integer quantum number. 
There are four series: (a) Finite dimensional nonunitary representations.
(b) Positive discrete series of infinite unitary representations, $D_j^+$. 
(c) Negative discrete series, $D_j^-$ and (d) Continuous series of unitary,
representations, $C_j$.\\
When we consider rotations in three dimensional Euclidean space, three
Euler angles are introduced for any arbitrary rotation, since the rotation
group possesses three compact generators. Similarly, in the case of $SU(1,1)$
one introduces three 'angles'. We can write a matrix element as $<j,m'|
e^{{iJ_3}{\cal{ \phi}}_1}e^{{iJ_2}{\eta}}e^{{iJ_3}{{\cal \phi}}_2}|jm>$, 
where ${{\cal \phi }}_1,~ {\eta},~ {{\cal \phi }}_2$ are the Euler angles.
The infinite dimensional representations are given below:\\
\bea
D_{mn}^{j_+}(a)={\Theta}_{mn}(j){\alpha ^*}^{-(m+n)}b^{m-n} {_2F_1}(-n-j,
1-n+j|1+m-n|-|b|^2)
\eea
with the constraint $m,n\ge-j>0,~~m\ge n$
\bea
D_{mn}^{j_+}(a)=(-1)^{n-m}{\Theta}_{nm}(j){\alpha }^{m+n}{b^{\star}}^{m-n} 
{_2F_1}(-m-j,
1-m+j|1+n-m|-|b|^2)
\eea                                                                            
with the condition that $m,n \ge -j>0,~~n\ge m$
\bea
{\Theta}_{mn}(j)= {1\over {(m-n)!}} \left[ {{(m+j)!(m-j-1)!}\over {(n+j)!
(n-j-1)!}} \right]^{1\over 2}
\eea
\bea
D_{mn}^{j_-}(a)={\Theta}_{mn}(j){\alpha ^*}^{m+n}b^{m-n} {_2F_1}(-m-j,
1+m+j|1+m-n|-|b|^2)
\eea
Now the constraints, for the negative discrete series are $m,n<j<0,~~m\ge n$
\bea
D_{mn}^{j_-}(a)=(-1)^{n-m}{\Theta}_{nm}(j){\alpha }^{m+n}{b^{\star}}^{m-n}
{_2F_1}(n-j,
1+n+j|1+n-m|-|b|^2)
\eea                                                                            
and the conditions to be satisfied are $m,n<j<0,~~n\ge m$
\bea
{\Theta}_{mn}(j)= {1\over {(m-n)!}} \left[ {{(-n+j)!(-n-j-1)!}\over {(-m+j)!
(-m-j-1)!}} \right]^{1\over 2}
\eea                                                                            
for the $D_{mn}^{j_-}$ series. The parameters appearing in these functions are
defined below,
\be
 j=-{k\over 2},~~~k=1,2,3....
\ee
\be
\alpha=e^{-i{\cal\phi}_1}{\rm cosh}
{{{\eta}\over 2}}e^{-i{\cal\phi }_2},~~{\rm and}~~
b=e^{-i{\cal\phi}_1}{\rm sinh}
{{{\eta}\over 2}}e^{i{\phi }_2}                        
\ee 
$a$ appearing in the argument collectively stands for three Euler angles.
The Casimir is $-j(j+1),~ j<0$.\\
Now we define the relevant C-functions for the continuous series
\bea
C_{mn}^j(a)={\Theta}_{nm}(j){\alpha ^{\star}}^{m+n}{b}^{m-n}
{_2F_1}({1\over 2}+m+\sigma,
{1\over 2}+m-{\sigma}|1+m-n|-|b|^2)
\eea                                                                            
and $m\ge n$
\bea
C_{mn}^j(a)={\Theta}_{nm}(j){\alpha }^{m+n}{b^{\star}}^{m-n}
{_2F_1}({1\over 2}+n+\sigma,
{1\over 2}+n-{\sigma}|1+n-m|-|b|^2)
\eea                                                                            
here $n\ge m$
\bea
{\Theta}_{mn}(j)= {1\over {(m-n)!}} {\Pi _{k=1}^{m-n}}
\left[ {1\over 4}-{\sigma}^2 +(n+k)(n+k-1) \right]^{1\over 2}, ~~m>n
\eea                                                                            
and
\bea
{\Theta}_{mn}(j)= {{(-1)^{n-m}}\over {(m-n)!}} {\Pi _{k=1}^{n-m}}
\left[ {1\over 4}-{\sigma}^2 +(m+k)(m+k-1) \right]^{1\over 2}, ~~n>m
\eea                                                                            
The Casimir $j(j+1)\equiv {1\over 4} -{\sigma}^2 = {1\over 4}+s^2=q$. For the
continuous series: ${\rm Re}~ j=-{1\over 2}, ~~0<~{\rm Im}~j~<\infty $ or
${\rm Im}~j=0, ~~-{1\over 2}~<{\rm Re}~j~<0 $
The orthogonality relation for the discrete series is
\bea
\label{ortho1}
\int da~D_{mn}^{j_{\pm}}(a){D_{m'n'}^{\star}}^{j'_{\pm}}=-{{\delta _{mm'}
\delta _{nn'}\delta _{jj'}}\over {2j+1}}
\eea
Moreover,
\bea
\int da~D_{mn}^{j_{\pm}}(a){D_{m'n'}^{\star}}^{j'_{\mp}} =0
\eea
For the continuous series, ${\rm Re}~j=-{1\over 2}$ and $o<{\rm Im}~j<\infty$,
the relevant orthonormality relation is
\bea\label{ortho2}
\int da T_{mn}^q(a){T_{m'n'}^{q'}}^*(a)={{2\pi\delta _{mm'}\delta _{nn'}
\delta (s-s')|\Gamma (2is) |^2}\over{|\Gamma ({1\over 2}-n-is)\Gamma 
({1\over 2} +n+is) |^2}} \eea
The  measure of the integration is given by
\bea
\label{measure}
\int ~da={1\over {2({2\pi})^2}}\int _0^{2\pi}d{\cal \phi}_1\int _0^{2\pi}
d{\cal{\phi}} _2\int _0^{\infty} d{\eta}~{\rm sinh}~{\eta}
\eea
It follows from a fundamental theorem in harmonic analysis with $SU(1,1)$
that we can bring together the two types of solutions: discrete and
continuous series discussed above. The completeness property can be used
to expand any square integrable function of the rotation angle, $\eta$,
in terms of the continuous series and finite number of representations of
the discrete series (see R\"uhl for example \cite{su11}).

\end{document}